\documentclass[apjl]{emulateapj}
\usepackage{amsfonts,amsmath,graphicx,natbib,apjfonts,subfigure}

\def\swift{{\it Swift}}
\def\chandra{{\it Chandra}}
\def\cfa{1}
\def\fer{2}
\def\col{3}
\def\mic{4}
\def\india{5}
\shorttitle{SN\,2012ap in the X-rays}
\shortauthors{Margutti et al.}

\begin{document}
\title{Relativistic supernovae have shorter-lived central engines or  more extended progenitors: the case of SN\,2012ap}

\author{R. Margutti\altaffilmark{\cfa}, D. Milisavljevic\altaffilmark{\cfa}, A.~M.~Soderberg\altaffilmark{\cfa}, 
C. Guidorzi\altaffilmark{\fer}, B.~J. Morsony\altaffilmark{\col}, N. Sanders\altaffilmark{\cfa}, S. Chakraborti\altaffilmark{\cfa}, 
A. Ray\altaffilmark{\india}, A. Kamble\altaffilmark{\cfa}, M. Drout\altaffilmark{\cfa}, J. Parrent\altaffilmark{\cfa}, A. Zauderer
\altaffilmark{\cfa}, L. Chomiuk\altaffilmark{\mic}}

\altaffiltext{\cfa}{Harvard-Smithsonian Center for Astrophysics, 60 Garden St., Cambridge, MA 02138, USA}
\altaffiltext{\fer}{Department of Physics and Earth Sciences, University of Ferrara, via Saragat 1, I-44122, Ferrara, Italy}
\altaffiltext{\col}{Department of Astronomy, University of Wisconsin-Madison, 2535 Sterling Hall, 475 N. Charter Street, Madison WI 53706
-1582, USA}
\altaffiltext{\mic}{Department of Physics and Astronomy, Michigan State University, East Lansing, MI 48824, USA}
\altaffiltext{\india}{Department of Astronomy and Astrophysics, Tata Institute of Fundamental Research, 1 Homi Bhabha Road, Mumbai 400 005, India}

\begin{abstract}
Deep late-time X-ray observations of the relativistic, engine-driven, type Ic SN\,2012ap allow us
to probe the nearby environment of the explosion and reveal the unique properties of relativistic SNe.
We find that on a local scale of $\sim0.01\,\rm{pc}$ the environment was shaped 
directly by the evolution of the progenitor star
with a pre-explosion mass-loss rate $\dot M <5 \times 10^{-6}\,\rm{M_{\sun}\,yr^{-1}}$, 
in line with GRBs and the other relativistic SN2009bb.
Like sub-energetic GRBs, SN\,2012ap is characterized by a bright radio emission and 
evidence for mildly relativistic ejecta. However, its late time ($\delta t\approx 20\,\rm{d}$) 
X-ray emission is $\sim100$ times fainter than the faintest sub-energetic GRB at the same epoch,
with no evidence for late-time central engine activity.
These results support  theoretical proposals that link relativistic SNe like 2009bb and 2012ap with the
 weakest observed engine-driven explosions, 
where the jet barely fails to breakout. Furthermore, our observations demonstrate that the difference
between relativistic SNe and sub-energetic GRBs is intrinsic and not due to line-of-sight
effects. This phenomenology can either be due to an intrinsically shorter-lived engine or to
a more extended progenitor in relativistic SNe.
\end{abstract}

\section{Introduction}
\label{Sec:Intro}

The vast majority of supernova explosions (SNe) arising from hydrogen and helium stripped progenitors
(i.e. type Ic SNe, see \citealt{Filippenko97} for the spectral classification of SNe) 
can be explained by the hydrodynamical collapse of the massive progenitor star (e.g. \citealt{Tan01}). 
In a very limited percentage of cases ($\lesssim 1$\%, \citealt{Berger03a,Coward05},
 \citealt{Guetta07}, \citealt{Soderberg10}), the explosion is
instead powered by an engine able to accelerate a tiny portion of the ejecta with typical mass\footnote{For 
the relativistic SN\,2009bb  $M\approx 10^{-2}\,\rm{M_{\sun}}$ \citep{Chakraborti11}.}
$M\approx 10^{-5}-10^{-6}\,M_{\sun}$ to velocities $v\gtrsim0.6\,c$.
Engine-driven explosions
(E-SNe hereafter) are thus uncommon. Furthermore, only a small fraction
of E-SNe ( $\lesssim10\%$, e.g. \citealt{Soderberg06}, see also 
\citealt{Cobb06,Pian06, Liang07,Virgili09}) harbor a fully 
relativistic jet and give origin to Gamma-Ray Bursts (GRBs).
The peculiar circumstances that cause a hydrogen-stripped, massive progenitor star
to produce a relativistic jet at the time of the collapse are still not fully understood. High
angular momentum seems to be a key ingredient (e.g. \citealt{MacFadyen99},
\citealt{MacFadyen01}, \citealt{Woosley06}, \citealt{Dessart08}).

E-SNe have historically been detected through their prompt X-ray and $\gamma$-ray emission 
produced by energy dissipation within the jet (ordinary GRBs) and by the SN shock break out 
(which is relevant at least for some sub-energetic -sub-E- GRBs, e.g.
\citealt{Kulkarni98,Matzner99,MacFadyen01,Tan01,Campana06, 
Wang07,Waxman07,Katz10,Bromberg11,Nakar12}). 
More recently, two E-SNe have been discovered through their bright
later-time radio emission (i.e. the relativistic SNe 2009bb and 2012ap, \citealt{Soderberg10,Bietenholz10,Chakraborti11b};
\citealt{Chakraborti14}, hereafter C14). Here we specifically ask the questions: 
what is the nature of relativistic SNe and what is their connection with the other
classes of E-SNe known so far?


The two known relativistic SNe 2009bb and 2012ap share with sub-E GRBs  evidence for
mildly relativistic ejecta powering a bright radio emission  
(\citealt{Soderberg06,Soderberg10,Bietenholz10}, C14) and a very energetic optical explosion with 
$E_{\rm{k}}\sim10^{52}\,\rm{erg}$ coupled to material moving at $v\sim$ a few $10^4\,\rm{km\,s^{-1}}$
(\citealt{Pignata11,Milisavljevic14b}). 
These properties dynamically distinguish relativistic SNe and sub-E GRBs from 
ordinary SNe, and put these explosions between the highly-relativistic, collimated GRBs 
and the more common type Ic SNe.

On the theoretical side, state-of-the art simulations of jet-driven stellar explosions (e.g. \citealt{Lazzati12})
associate classic GRBs with fully-developed, highly-relativistic jets and suggest that 
sub-E GRBs likely represent the cases where the jet is just barely able to pierce 
through the stellar envelope (\citealt{Bromberg11}, \citealt{Nakar12}). In particular, it
was suggested by \citealt{Lazzati12} that a different life-time of the central engine
might be able to explain the entire zoo of E-SNe (i.e. relativistic SNe, sub-E GRBs
and ordinary GRBs). However, it is unclear 
if relativistic SNe represent a new class of explosions, or if, instead, they are the equivalent
of sub-E GRBs for which we missed the high-energy trigger because of
line-of-sight effects or incomplete coverage of the $\gamma$-ray satellites
(see e.g. the discussion for SN\,2009bb in \citealt{Soderberg10}). This still-open-question
motivates this study.

We present late-time deep X-ray observations of the relativistic
SN\,2012ap. These observations allow us to 
identify for the first time a distinctive property of relativistic SNe
that clearly sets them apart from all the other known engine-driven explosions.
We find that relativistic SNe are characterized by a significantly fainter X-ray emission
at late times ($t\sim 20$ d), even compared to sub-E GRBs (SN\,2012ap is $\sim100$ times
fainter than the faintest sub-E GRB at the same epoch), and show no evidence for
an excess of X-ray radiation that has been recently reported for the sub-E
GRBs 060218 (\citealt{Soderberg06, Fan06}) and 100316D (\citealt{Margutti13b}).

We describe our observations in Sec. \ref{Sec:Obs} and constrain the progenitor mass-loss
rate in Sec. \ref{Sec:massloss}. Finally we put SN\,2012ap in the context of engine-driven 
explosions in Sec. \ref{Sec:CE} and \ref{Sec:discussion} and discuss how our findings
clearly suggest that relativistic SNe constitute a separate class of engine-driven
explosions with intrinsic differences with respect to sub-E GRBs.
Conclusions are drawn in Sec. \ref{Sec:conclusions}.

Uncertainties are quoted at $1\sigma$ confidence level, unless otherwise noted.
We employ standard cosmology with $H_{0}=71$ km s$^{-1}$ Mpc$^{-1}$,
$\Omega_{\Lambda}=0.73$, and $\Omega_{\rm M}=0.27$.
Throughout the paper we use 2012 February 5th as the explosion date of 
SN\,2012ap, as inferred by \cite{Milisavljevic14b} (M14, hereafter) from extensive
optical observations. Following  C14
we assume a distance of 40 Mpc \citep{Springob07,Springob09}. 
A detailed discussion of the optical and radio properties of SN\,2012ap
can be found in M14 and C14, respectively.
Finally we note that sub-E GRBs are also called low-luminosity GRBs in the
literature (see e.g. \citealt{Bromberg11}). However the physical parameter that is 
relevant to your analysis is the (modest) kinetic energy of their fastest ejecta,
which is surely related to the low-luminosity of their prompt $\gamma$-ray emission. 
For this reason we will refer to this class as sub-E GRBs.
\section{Observations and data analysis}
\label{Sec:Obs}

\subsection{\swift-XRT}
\label{SubSec:XRTObs}

We observed SN\,2012ap with the \emph{Swift} \citep{Gehrels04} 
X-ray Telescope (XRT, \citealt{Burrows05}) starting from 2012 February 12th
($\delta t\approx 7\,\rm{d}$) until March 2nd  ($\delta t\approx26\,\rm{d}$). 
No X-ray source is detected at the position of SN\,2012ap.
Analyzing the XRT data using the latest HEAsoft release (v6.13) and employing
standard filtering  and screening criteria, we determine a $3\,\sigma$ 
count-rate upper limit to the X-ray emission from SN\,2012ap of $7.3\times 
10^{-4}\,\rm{cps}$ (0.3-10 keV energy band, total exposure time of 35 ks).
The Galactic neutral hydrogen column density in the direction of SN\,2012ap is 
$\rm{NH}=4.9\times 10^{20}\,\rm{cm^{-2}}$ \citep{Kalberla05}. The analysis of the
optical spectra presented in \cite{Milisavljevic14} constrains the intrinsic
color excess towards SN\,2012ap to be $0.18\,\rm{mag}<E(B-V)<0.57\,\rm{mag}$.
Using the Galactic relations between the extinction $A_V$ and the $\rm{NH}$
($\rm{NH}/A_V\approx (1.7-2.2)\times 10^{21}\,\rm{cm^{-2}}$, \citealt{Predehl95},
\citealt{Watson11}), the limit on the color excess above translates into 
an intrinsic neutral hydrogen column density
$\rm{NH}_{\rm{x,i}}<3.9\times 10^{21}\,\rm{cm^{-2}}$.
Assuming a simple power-law spectral model with photon index $\Gamma=2$, the
absorbed (unabsorbed) flux limit is $F_{\rm{x}}<2.6\times 10^{-14}\,\rm{erg\,s^{-1}cm^{-2}}$
($F_{\rm{x}}<5.8\times 10^{-14}\,\rm{erg\,s^{-1}cm^{-2}}$), corresponding to a 
luminosity $L_{\rm{x}}<1.1\times 10^{40}\,\rm{erg\,s^{-1}}$ (0.3-10 keV) at the distance
of $40\,\rm{Mpc}$. 
\subsection{\chandra}
\label{SubSec:ChandraObs} 

\begin{figure}
\vskip -0.0 true cm
\centering
\includegraphics[scale=0.385]{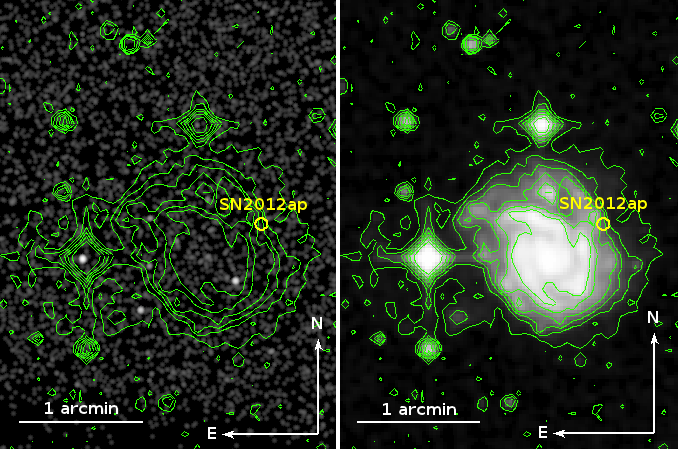}
\caption{X-ray (\emph{Chandra}, 0.5-8 keV, left panel) and pre-explosion optical 
image (SDSS, right panel) of the region around SN\,2012ap. No X-ray emission
is detected at the position of SN\,2012ap at $\delta t\approx 24\,\rm{d}$ after the
explosion down to a deep luminosity
limit of $L_{\rm{x}}\sim2\times 10^{39}\,\rm{erg\,s^{-1}}$ (0.3-10 keV).
Yellow circle: $2\arcsec$ radius region around SN\,2012ap. Optical contours 
have been overlaid to the X-ray image for reference.}
\label{Fig:Chandra}
\end{figure}

We initiated deep X-ray follow up of SN\,2012ap with the \emph{Chandra} X-ray
Observatory on 2012 Feb 29.2 UT, $\delta t\approx 24\,\rm{d}$
after the explosion (Program 13500648; PI Soderberg).
\emph{Chandra} ACIS-S data were reduced with the {\tt CIAO} software package
(v4.5) and relative calibration files, applying standard ACIS data filtering. 
Using {\tt wavedetect} we find no evidence for X-ray emission at the position of
SN\,2012ap (Fig. \ref{Fig:Chandra}), with a $3\sigma$ limit of $8.0\times 10^{-4}\rm{cps}$ (0.5-8 keV
energy range, total exposure time of $9.9$ ks). Employing the spectral parameters
above, the corresponding absorbed (unabsorbed) flux limit in the 0.3-10 keV
energy range is $F_{\rm{x}}<6.8\times 10^{-15}\,\rm{erg\,s^{-1}cm^{-2}}$ 
($F_{\rm{x}}<1.3\times 10^{-14}\,\rm{erg\,s^{-1}cm^{-2}}$). The luminosity limit 
is $L_{\rm{x}}<2.4\times 10^{39}\,\rm{erg\,s^{-1}}$ (0.3-10 keV).
\section{Constraints on the progenitor mass-loss rate}
\label{Sec:massloss}

\begin{figure*}
\vskip -0.0 true cm
\centering
\includegraphics[scale=0.7]{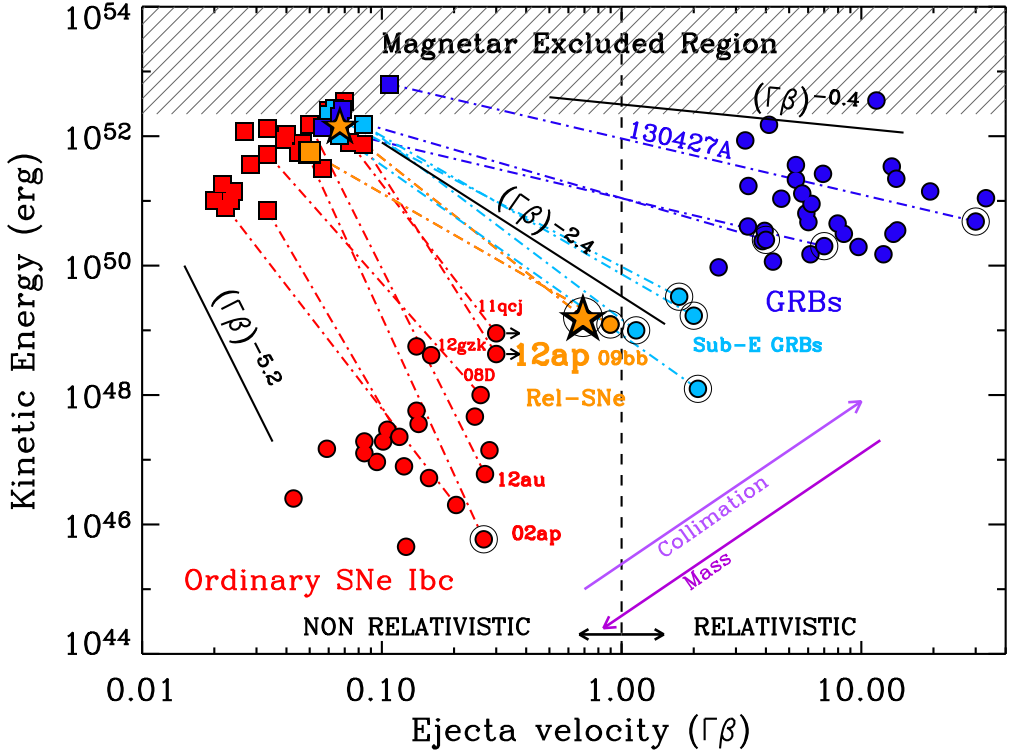}
\caption{Kinetic energy profile of the ejecta of ordinary type Ibc SNe (red) and
E-SNe, a class of explosions that  includes GRBs (blue),
sub-E GRBs (light-blue) and relativistic SNe (orange). Squares and circles are used
for the slow-moving and the fast-moving ejecta, respectively, as measured from optical
and radio observations. The velocity of the fast-moving ejecta has been computed at
$\delta t=1\,\rm{d}$ (rest-frame). Black solid lines: ejecta kinetic energy
profile of a pure hydrodynamical explosion ($E_{\rm{k}}\propto (\Gamma \beta)^{-5.2}$,
\citealt{Tan01}), and for explosions powered by a short-lived ($E_{\rm{k}}\propto (\Gamma \beta)^{-2.4}$)
and long-lived ($E_{\rm{k}}\propto (\Gamma \beta)^{-0.4}$) central engine (\citealt{Lazzati12}). 
Open black circles identify explosions with broad-lined
optical spectra. The purple arrows identify the directions of increasing 
collimation and mass of the fastest ejecta.
SN\,2012ap bridges the gap between cosmological GRBs and 
ordinary SNe Ibc. Its kinetic energy profile, significantly flatter than what expected from 
a pure hydrodynamical explosion, indicates the presence of a central engine. References:
\cite{Margutti13b} and references therein; \cite{Ben-Ami12}; \cite{Horesh13}; \cite{Corsi14},
\cite{Walker14}; C14; M14.}
\label{Fig:EkGammaBeta}
\end{figure*}

At $\delta t\lesssim30\,\rm{d}$ Inverse Compton (IC) is the dominating X-ray emission mechanisms
for ordinary SNe arising from hydrogen-stripped progenitors exploding in low density environments
(\citealt{Bjornsson04}; \citealt{Chevalier06}). In the case of central-engine powered SNe, additional
sources of X-ray power are represented by continued central engine activity (as in the case of sub-E
GRBs like 100316D, \citealt{Margutti13b}) and interaction of the explosion jet with the environment (as
in the case of ordinary GRBs, see e.g. \citealt{Margutti13}). In the following we use the deep \emph{Chandra}
limit of Sec. \ref{SubSec:ChandraObs} and conservatively assume that IC is responsible for the entire 
X-ray emission to derive a solid upper limit to the mass-loss rate of the progenitor star of SN\,2012ap.

In the IC scenario the X-ray emission is originated by up-scattering of optical photons from the 
SN photosphere by a population of relativistic electrons and depends on: (i) the density structure of the
SN ejecta (ii) and of the circum-stellar medium (CSM); (iii) the details of the electron distribution 
responsible for the up-scattering; (iv) the explosion parameters (ejecta mass $M_{\rm{ej}}$ and 
kinetic energy\footnote{This is the kinetic energy carried by the slowly moving material powering the
optical emission.} $E_{\rm{k}}$); and (v) the bolometric luminosity of the SN: $L_{\rm{IC}}\propto L_{\rm{bol}}$.
We adopt the formalism by \cite{Margutti12} modified to account for the outer density structure of 
SNe with compact progenitors that has been shown to scale as $\rho_{\rm{SN}}\propto R^{-n}$ with $n\sim10$ 
(see e.g. \citealt{Matzner99}; \citealt{Chevalier06}). 

Assuming a wind-like CSM structure $\rho_{\rm{CSM}}\propto
R^{-2}$ as appropriate for massive stars, a power-law electron distribution $n_{e}(\gamma)=n_0 \gamma^{-p}$ with 
$p\sim3$ as indicated by radio observations of type Ib/c SNe \citep{Chevalier06} and by radio observations 
of SN\,2012ap (C14) and a fraction of energy into relativistic electrons 
$\epsilon_e=0.1$ as supported by well studied SN shocks (e.g.  \citealt{Chevalier06}), the \emph{Chandra} non-detection
of SN\,2012ap at $\delta t\approx 24\,\rm{d}$ implies $\dot M/v_w <5\times10^{-6}(M_{\sun}y^{-1}/1000\rm{km\,s^{-1}})$.
$\dot M$ is the mass loss rate of the progenitor star and $v_w$ is the wind velocity. We renormalize the
mass-loss to $v_w=1000\,\rm{km\,s^{-1}}$ as appropriate for a Wolf Rayet progenitor stars. In this calculation we 
used the bolometric luminosity we derived in M14, 
$E_{\rm{k}}\sim 10^{52}\,\rm{erg}$ and $M_{\rm{ej}}\sim 3\,\rm{M_{\sun}}$ as obtained by modeling the bolometric
luminosity in M14. 

The inferred limit to the mass-loss rate $\dot M <5\times10^{-6}(M_{\sun}y^{-1})$  
is independent from any assumption on magnetic-field related parameters, it is not affected 
by possible uncertainties on the SN distance and indicates that the pre-explosion mass-loss
of SN\,2012ap lies at the low end of the interval of values derived by C14
($4\times10^{-6}\,\rm{M_{\sun}y^{-1}}<\dot M <5\times10^{-5}\,\rm{M_{\sun}y^{-1}}$) based on the 
modeling of the radio observations with synchrotron emission.\footnote{Note that the synchrotron
formalism is instead dependent on assumptions on magnetic field related parameters.}
This result is in line with the value derived for the relativistic SN\,2009bb ($\dot M\sim2\times10^{-6}
\,\rm{M_{\sun}y^{-1}}$, \citealt{Soderberg10})
and consistent with the wide range of values inferred for sub-E GRBs ($10^{-7}\,\rm{M_{\sun}y^{-1}}\lesssim\dot M 
\lesssim10^{-5}\,\rm{M_{\sun}y^{-1}}$).

\begin{figure*}
\vskip -0.0 true cm
\centering
\includegraphics[scale=1]{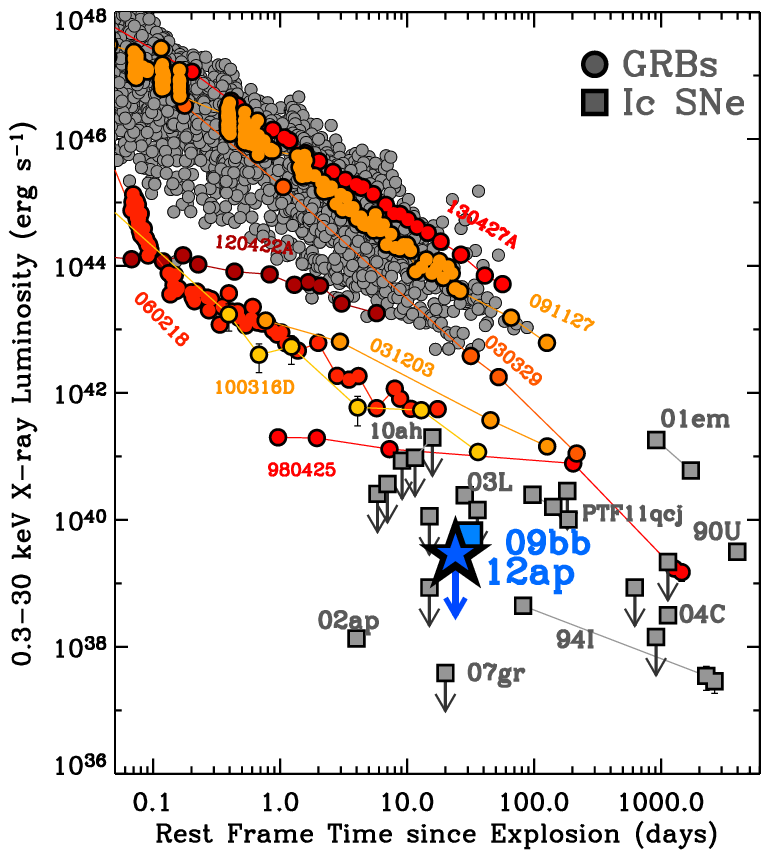} 
\includegraphics[scale=0.6]{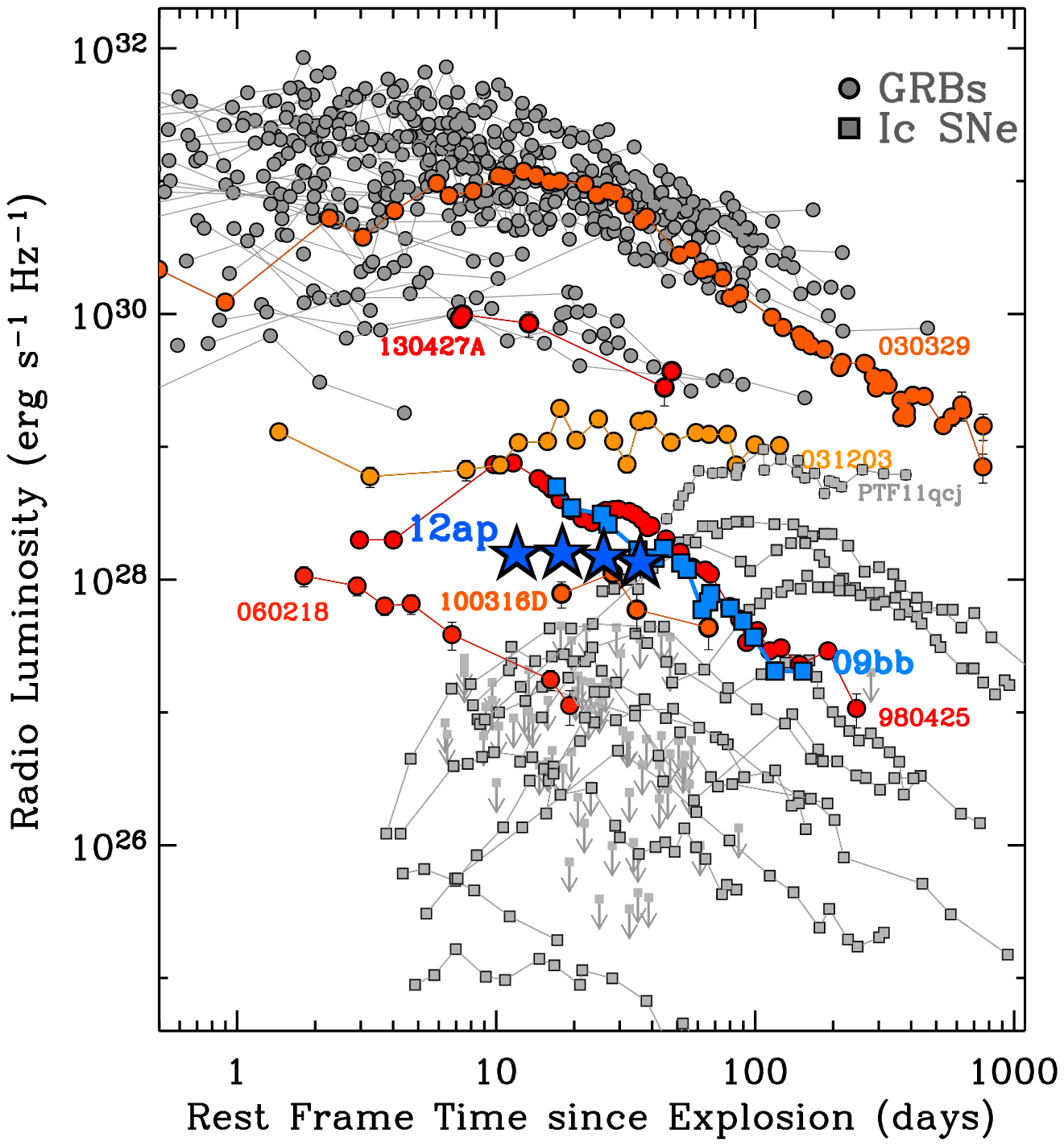}
\caption{\emph{Left panel}: \emph{Chandra} observations put a deep limit to the X-ray luminosity of the
relativistic SN\,2012ap at $\sim20$ days after the explosion. 
SN\,2012ap is considerably less luminous than ordinary long GRBs (filled circles, from \citealt{Margutti13b},
\citealt{Margutti13} and referenced therein) and  is $\sim100$ times fainter than 
the faintest sub-E GRBs (i.e. GRBs 980425 and 100316D). 
Filled grey squares: X-ray emission from ordinary type Ic SNe. The relativistic SN2009bb
is marked with a blue square. References: \cite{Immler02}, \cite{Pooley04}, \cite{Soria04},
\cite{Soderberg05}, \cite{Perna08}, \cite{Corsi11}, \cite{Horesh13}, \cite{Corsi14}.
\emph{Right panel}: radio emission of SN\,2012ap (from 
C14) compared to a sample of GRB radio afterglows (filled circles) and
type Ic SNe (filled square) collected from \cite{Soderberg10}, \cite{Corsi11}, \cite{Chandra12}, \cite{Horesh13}, \cite{Margutti13b} and cite{Corsi14}. 
At radio frequencies the luminosity of SN\,2012ap is comparable to (or even larger than) sub-E GRBs.
In both panels GRBs with spectroscopically associated SNe are in color and labeled. Different
shades of orange and red are used to guide the eye.}
\label{Fig:Xray}
\end{figure*}

\begin{figure}
\vskip -0.0 true cm
\centering
\includegraphics[scale=0.44]{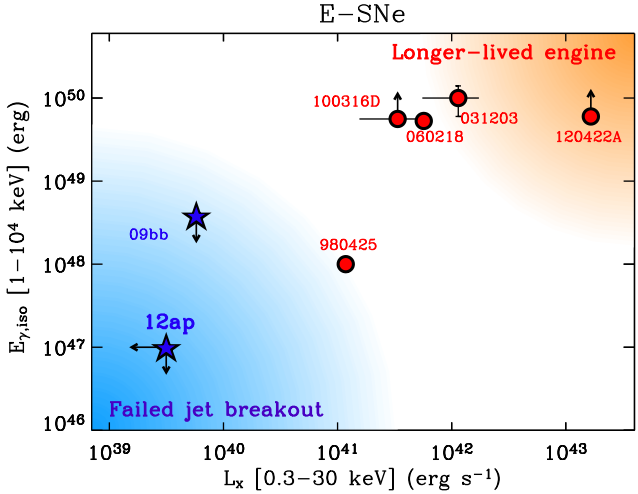} 
\caption{Promptly emitted $\gamma$-ray energy vs. X-ray luminosity between 10 and 30 days
since the explosion for the sample of relativistic SNe (blue stars) and sub-E GRBs (red circles).
Relativistic SNe are clearly distinguished from sub-E GRBs by their significantly 
fainter X-ray emission. References: \cite{Amati06}; \cite{Soderberg06c}; \cite{Soderberg10}; \cite{Starling11}
\cite{Barthelmy12}; \cite{Margutti13b}; \cite{Margutti13}; \cite{Amati13}; \cite{Amati13b}; C14.}
\label{Fig:LxEiso}
\end{figure}

\begin{figure}
\vskip -0.0 true cm
\centering
\includegraphics[scale=0.43]{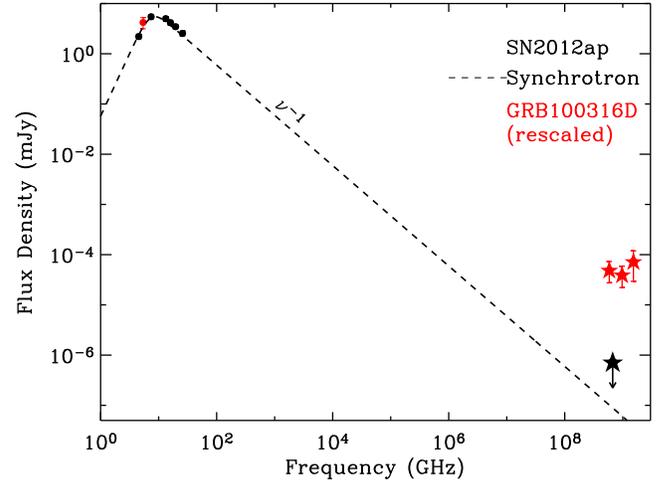} 
\caption{Radio (filled black circles) to X-ray (black stars) SED of SN\,2012ap. The \emph{Chandra} X-ray upper limit
is consistent with the extrapolation of the best-fitting synchrotron model obtained by 
C14 at $\delta t\approx 20$ d. Notably, the X-ray emission from 
SN\,2012ap is $\geq 100$ times fainter than the sub-E GRB\,100316D at a similar epoch (here
rescaled to match the level of the detected SN\,2012ap radio emission), thus ruling out the presence of
an extra X-ray component arising from the activity of the explosion central engine.}
\label{Fig:Radio2Xray}
\end{figure}

\section{SN\,2012ap in the context of engine-driven explosions}
\label{Sec:CE}

The radio observations of SN\,2012ap are well modeled by synchrotron emission
arising from the interaction of the SN shock with the environment (C14). 
C14 derive $E_{k}=(1.6\pm0.1)\times 10^{49}\,\rm{erg}$ carried 
by mildly relativistic ejecta
with velocity $v\sim0.7c$ at $\delta t=1\,\rm{d}$. By modeling the observed optical emission,
M14 infer $E_{\rm{k}}\sim10^{52}\,\rm{erg}$ in slow moving 
($v\approx 20000\,\rm{km\,s^{-1}}$) material. These two values define 
an $E_{\rm{k}}$ profile 
significantly flatter than what expected in the case of a pure hydrodynamical 
collapse ($E_{\rm{k}}\propto (\Gamma \beta)^{-5.2}$, e.g. \citealt{Tan01}), 
thus pointing to the \emph{presence of an engine driving the
SN\,2012ap explosion} (see Fig. \ref{Fig:EkGammaBeta}). 

Engine-driven SNe (E-SNe)
constitute a diverse class of explosions that includes relativistic SNe, sub-E GRBs and ordinary GRBs.
SN\,2012ap is intermediate between ordinary non-relativistic SNe and fully relativistic GRBs and falls into a
region of the parameter space populated by sub-E GRBs and the other known relativistic SN, 
SN\,2009bb (Fig. \ref{Fig:EkGammaBeta})\footnote{The relativistic nature of SN\,2007gr
has been questioned by \cite{Soderberg10b} and it is not included here. See however \cite{Paragi10}.}. 
With reference to figures \ref{Fig:Xray} and \ref{Fig:LxEiso} we find that:
\begin{itemize}
\item The radio luminosity of SN\,2012ap and sub-E GRBs is comparable. 
SN\,2012ap is significantly more luminous than ordinary Ic SNe
at the same epoch, and even more luminous than the sub-E GRBs 100316D and
060218  (Fig. \ref{Fig:Xray}, right panel). With $E_{\rm{k}}\sim 10^{52}\,\rm{erg}$ and evidence
for broad spectral features (M14), the properties of 
SN\,2012ap in the optical band are also reminiscent of the very energetic SNe associated 
with sub-E GRBs and ordinary GRBs.
\item At $\delta t\sim 20\,\rm{d}$, the X-ray emission from SN\,2012ap is however a 
factor $\ge 100$ fainter then the faintest sub-E GRB ever detected, GRB\,980425
(Fig. \ref{Fig:Xray}, left panel). 
\item Along the same line, from C14,
the prompt $\gamma$-ray energy released by the SN\,2012ap explosion is $E_{\rm{\gamma,iso}}<
10^{47}\,\rm{erg}$, a factor $\ge 10$ fainter then the faintest sub-E GRB\,980425 (Fig. \ref{Fig:LxEiso}).
\end{itemize}

In addition, in \cite{Milisavljevic14} and M14 we showed that:
\begin{itemize}
\item Contrary to sub-E GRBs and GRBs, SN\,2012ap exploded in a solar-metallicity environment.
Interestingly, the metallicity of the environment of SN 2009bb was also super-solar \citep{Levesque10}.
\item Differently from sub-E GRBs and GRBs, our 
analysis of multi-epoch spectroscopy strongly favors the presence of helium in the
ejecta of SN\,2012ap.  Helium was also reported in the early-time spectra of SN\,2009bb
\citep{Pignata11}.
\end{itemize}

\emph{Relativistic SNe and sub-E GRBs are thus clearly distinguished in terms of their high-energy 
(X-rays and $\gamma$-rays) properties, a higher metallicity environment and the
conspicuous presence of helium in their ejecta.}

The different level of X-ray emission between relativistic SNe and sub-E GRBs cannot be 
ascribed to beaming of collimated emission away from our line of sight.
Radio observations of sub-E GRBs support the idea
of quasi-spherical explosions (e.g. \citealt{Kulkarni98}, \citealt{Soderberg04}, \citealt{Soderberg06}, 
\citealt{Margutti13b}), and there is no
evidence for beaming of the non-thermal emission from relativistic SNe 
(\citealt{Soderberg10}; C14).
Furthermore, on a time scale of $\sim 20\,\rm{d}$, the blastwave arising from both 
relativistic SNe and sub-E GRBs is sub-relativistic and the geometry of emission is 
effectively spherical, independent from the initial conditions. The different level of
X-ray emission between sub-E GRBs and relativistic SNe at $t\gtrsim 10\,\rm{d}$ is thus intrinsic.

While both relativistic SNe and sub-E GRBs are  intermediate between ordinary type Ic SNe and 
GRBs, these findings point to a diversity in the properties of the progenitors and/or 
the engines that drive their explosion. This topic is discussed below.
\section{Discussion}
\label{Sec:discussion}

At $\delta t\gtrsim 10\,\rm{d}$ the detected X-ray emission
from sub-E GRBs like  060218, 100316D has been suggested to originate from
the activity of the explosion central engine (\citealt{Soderberg06}, \citealt{Fan06}, 
\citealt{Fan11}, \citealt{Margutti13b}), 
which dominates over synchrotron emission 
from the shock-CSM interaction.\footnote{As noted in \cite{Margutti13b}, this
extra component might be present in classical GRBs as well, but it is likely
out-shined by emission from the the jet-CSM interaction.} 
The nature of the central engine is currently not known. For the 
sub-E GRBs 060218 and 100316D the observations support either a magnetar central engine 
or continued accretion onto a newly formed black-hole.

Figure \ref{Fig:Radio2Xray} clearly shows that the X-ray emission from SN\,2012ap 
is instead consistent with the shock-CSM model that best fits the radio observations. 
For SN\,2012ap, the deep X-ray limit thus
rules out the presence of an additional, luminous X-ray component arising from the engine 
activity, contrary to sub-E GRBs like 100316D portrayed in Fig. \ref{Fig:Radio2Xray}. 
This finding suggests that the engine that powers SN\,2012ap is short lived and unable to survive 
for such a long time.

We propose that relativistic SNe like 2009bb and 2012ap represent weak engine-driven
explosions, where the engine activity stops before being able to produce a successful jet breakout.
The result is a stellar explosion that is able to accelerate a tiny fraction of ejecta to 
mildly relativistic velocities, thus dynamically different from ordinary Ic SNe and more similar
to sub-E and classical GRBs (Fig. \ref{Fig:EkGammaBeta}). 
In contrast to GRBs, however, the jet is not able to pierce through the stellar 
envelope, and a very limited fraction of energy is dissipated at 
$\gamma$-ray frequencies, consistent with the deep limit $E_{\rm{\gamma,iso}}<10^{47}\,\rm{erg}$
from C14. This phenomenology can either be due to an intrinsically short-lived engine
or to a different progenitor structure/properties between relativistic SNe and GRBs. 
We discuss these two possibilities in Sec. \ref{SubSec:CE}, \ref{SubSec:progenitor1}
and \ref{SubSec:progenitor2}.

\subsection{Central engine life-time}
\label{SubSec:CE}

\cite{Lazzati12} investigated the role of the duration of the engine activity in 
stellar explosion induced by relativistic jets with a set of numerical simulations.
These authors find that the duration of the engine activity is a key parameter
that determines the outcome of a stellar explosion. For a fixed energy budget
and progenitor structure, \cite{Lazzati12} show that the longest-lived engines
always produce a successful explosion with a fully relativistic jet (i.e. a classical
GRB). Engines with intermediate and short durations (where ``short'' or ``long'' is
here referred to the time it takes to the jet head to break out through the stellar
envelope) would lead instead to partially or totally failed jets, respectively. 
In particular, relativistic SNe would result from explosions where the engine turns off on 
the jet breakout time-scale ($5-10\,\rm{s}$ depending on the energy budget and progenitor,
\citealt{Morsony07}, \citealt{Lazzati12}, their Fig. 5), in agreement with our observational
findings of shorter-lived engines. 
\subsection{Progenitor properties: metallicity}
\label{SubSec:progenitor1}

A failed jet breakout in relativistic SNe can also be due to different
progenitor properties compared with sub-E GRBs and ordinary GRBs.  
To this respect it is important to note that (i) the relativistic SNe 2009bb and 2012ap
exploded in super-solar and solar \emph{metallicity} environments, respectively, in
line with ordinary Ic SNe (e.g. \citealt{Sanders12b, Kelly12} and reference therein) but
in sharp contrast with sub-E GRBs and classical GRBs that show a marked
preference for sub-solar metallicity environments (e.g. \citealt{Stanek06,Margutti07,Modjaz08,Levesque10b}). 
(ii) Evidence for helium-rich ejecta was found for both
SNe 2009bb \citep{Pignata11} and 2012ap (M14), which points to the 
presence of a helium layer at the time of the explosion (i.e. a non-entirely envelope-stripped
progenitor star).

The metallicity of the two relativistic SNe known so far is actually large
even compared to the sample of energetic broad-lined Ic SNe \emph{not}
connected to GRBs and sub-E GRBs (\citealt{Sanders12b, Kelly12}). 
For SN\,2009bb \cite{Levesque10} estimate $\rm{Z}=1.7-3.5\,\rm{Z_{\sun}}$,
while for SN\,2012ap \cite{Milisavljevic14} find $\rm{Z}=1.0\,\rm{Z_{\sun}}$,
where solar metallicity $\rm{Z_{\sun}}$ corresponds to $\rm{log(O/H)} + 12 = 8.69$
\citep{Asplund05}.  At such high metallicity, line-driven winds in massive stars 
(e.g. \citealt{Castor75}) more efficiently strip away angular momentum from the progenitor 
through a more sustained mass loss
($\dot M\propto Z^{0.86}$ in Wolf-Rayet stars, likely progenitors of GRBs, 
\citealt{Vink05}).\footnote{Recent findings indicate that episodic mass-loss
episodes, as opposed to steady mass-loss through winds, also
have a role in the evolution of massive stars. However, the metallicity dependence of
these episodes of explosive mass-loss has yet to be constrained. See \cite{Margutti14}
and references therein for details.}  High angular momentum of the progenitor at collapse has been identified by
recent numerical simulations (e.g. \citealt{MacFadyen99,MacFadyen01,Woosley06}) 
as a key physical ingredient of the ``collapsar''
model to explain the presence of fully relativistic jets in GRBs.
It is thus possible that the higher metallicity of the progenitors of relativistic SNe 
inhibited the formation of a powerful jet able to pierce through 
the stellar envelope. 

The growing sample of GRBs
discovered in high-metallicity environments (see e.g. GRBs 050826, 051022 
\citealt{Graham13}, their Fig. 3 and  GRB\,120422A, intermediate between ordinary GRBs and
sub-E GRBs, \citealt{Shulze14})  points however to a more complex situation, where
metallicity has some role, but it is unlikely to be the ultimate parameter
driving the distribution of angular momentum at collapse.\footnote{A similar conclusion 
is reached by studies of the close environment of GRBs and energetic broad-lined Ic SNe
in the local universe. See  \cite{Sanders12b, Kelly14}.}
These findings suggest
that the higher metallicity of the two known relativistic SNe compared to GRBs might \emph{not}  be 
directly linked to the final explosion outcome. While it might still indicate a preference 
for different environments,\footnote{To this respect it is intriguing 
to note that for both events the SN spectrum showed unusually strong
signs of interaction with carriers of diffuse interstellar bands (DIBs) as we detail in \cite{Milisavljevic14}. 
These observations suggest that the material responsible for the detected DIBs is 
local to the SN explosion, and  possibly related to the mass-loss of the progenitor star
in the decades to years before the terminal explosion. Alternatively, it could point to a peculiar 
small-scale environment in which the progenitors of relativistic SNe preferentially form.}
it is important to note that differently from GRBs and sub-E GRBs, the
relativistic SNe were discovered by surveys targeting high-mass and hence metal-rich galaxies.
At the time of writing it is not clear if the higher metallicity of relativistic
SNe is simply the result of this observational bias.
\subsection{Progenitor properties: helium-rich ejecta}
\label{SubSec:progenitor2}

The fate of a newly born jet in a massive star is also tightly related to the
size and structure of the progenitor star at the time of collapse, as 
the first requirement for a successful jet breakout is the ability to cross the progenitor
and pierce through its envelope (e.g. \citealt{MacFadyen99}, \citealt{MacFadyen01},
\citealt{Lazzati12}
and references therein). The detection of helium-rich ejecta 
in SNe 2009bb and 2012ap indicates a non-complete shredding of the outer helium layers
of their progenitors before exploding, as opposed to the stripped type Ic SNe
associated with GRBs and sub-E GRBs. It is thus possible that the jet failed because
it was dumped by the additional helium layers of the stellar progenitors of
relativistic SNe.

A similar scenario of a weak jet, dumped by the external
helium layers of the stellar progenitor was proposed for 
the type Ib SN\,2008D by \cite{Mazzali08} to explain
the large kinetic energy ($E_k>10^{51}\,\rm{erg}$), 
early disappearance of broad spectral features and the 
serendipitous detection of a powerful X-ray flash of radiation 
with $L_x\sim6\times10^{43}\,\rm{erg\,s^{-1}}$ signaling 
the onset of the explosion. Later time radio observations
pointed however to a modest velocity of the freely expanding fastest ejecta
($\beta\approx 0.25$, \citealt{Soderberg08}, \citealt{Bietenholz10}).
Together with the detection of strong helium lines and the disappearance 
of the broad spectral features,
this finding clearly sets SN\,2008D apart from 
relativistic SNe, sub-E GRBs and GRBs (see Fig. \ref{Fig:EkGammaBeta})
and offers the case for an alternative
explanation of the initial X-ray flash as shock breakout radiation from an ordinary 
SN \citep{Soderberg08}.

While the nature of the X-ray flash is still under debate (e.g. \citealt{Tanaka09}, 
\citealt{Modjaz09}, \citealt{vanderHorst11}, \citealt{Couch11}, \citealt{Bersten13},
\citealt{Svirski14}), SN\,2008D brought to light the possibility that even ordinary SNe
with a thicker helium envelope might be triggered by bipolar jets (\citealt{Mazzali08},
\citealt{Xu08}), thus pointing to a continuum of properties bridging ordinary explosions
and GRBs. 
\subsection{A continuum of stellar explosions originating from hydrogen-stripped progenitors}
\label{SubSec:continuum}

Figure \ref{Fig:EkGammaBeta} strongly argues in favor of a continuum 
of properties of the fastest ejecta of 
stellar explosions originating from hydrogen-stripped progenitors (i.e. type Ib/c SNe). 
This finding is not new (see e.g.  \citealt{Xu08}, \citealt{Mazzali08}) 
and might be the observational manifestation of a continuum 
of  properties of the jets that power these explosions and can potentially 
result from (i) different central engine life-times or (ii) progenitor properties 
(i.e. metallicity or degree of stripping of the external helium layers).

While the kinetic energy profile of the ejecta E-SNe (i.e. orange, light-blue and
blue dots in Fig. \ref{Fig:EkGammaBeta}) points to the presence of a
central engine that drives the explosion, it is possible that even ordinary type Ib/c
SNe (i.e. red dots of Fig. \ref{Fig:EkGammaBeta}) are triggered by failed 
bipolar jets -as opposed to the generally assumed neutrino deposition explosion
mechanisms- that would leave no detectable imprint on the dynamics 
of the ejecta (\citealt{Khokhlov99}, \citealt{Granot04}, \citealt{Wheeler10},
 \citealt{Lazzati12}, \citealt{Nagakura12}).

This theoretical suggestion has been paralleled by a growing number of
``transitional'' objects found by recent surveys. Particularly relevant in this 
respect is the class of type Ic SNe with broad features in their spectra (i.e. Ic-BL,
famous historical examples are SN\,1997ef and SN\,1997dq, \citealt{Mazzali00},
\citealt{Mazzali04}).
While \emph{all} E-SNe are Ic-BL, not every type Ic-BL SN showed unambiguous evidence
for a central engine (\citealt{Soderberg06}, \citealt{Soderberg10}). 
Notable examples include the type Ic-BL SNe 2002ap (\citealt{Berger02}, \citealt{Gal-Yam02}, \citealt{Mazzali02}), 2010ay
(\citealt{Sanders12}), 2010ah (\citealt{Corsi11}, \citealt{Mazzali13}) and PTF10qts (\citealt{Walker14}).
However, persistent broad line features are found in association with large bulk kinetic energies of the 
ejecta ($E_k\sim10^{52}\,\rm{erg}$) and are indicative of large photospheric expansion velocities 
that might be powered by a jet that did not emerge from the progenitor.

Finally, the growing sample of ordinary type Ib/c SNe with larger-than-average velocities of their
fastest ejecta ($v\sim0.3\,c$ vs. $v\sim0.15\,c$, as it was recently found for PTF11qcj, SN\,2012au and
PTF12gzk, \citealt{Corsi14}, \citealt{Kamble13}, \citealt{Horesh13}) further strengthens the idea of a continuum
of hydrogen-stripped explosions, encompassing even ordinary SNe (Fig \ref{Fig:EkGammaBeta}).

\section{Summary and Conclusions}
\label{Sec:conclusions}
The class of engine-driven SNe (E-SNe) collects a rare variety of SN 
explosions ($\lesssim1\%$ of type Ic SNe) and includes relativistic SNe, sub-E GRBs and ordinary GRBs. 
E-SNe are characterized by a significantly shallower kinetic energy profile of the 
explosion ejecta than  expected in the case of a pure hydrodynamic
collapse of the progenitor star (Fig. \ref{Fig:EkGammaBeta}), indicating that E-SNe
are able to accelerate a tiny but important fraction of their ejecta to higher velocities ($v\gtrsim 0.6\,c$). 
E-SNe otherwise show a diverse phenomenology. 

Relativistic SNe and sub-E GRBs share a bright radio emission and evidence
for mildly relativistic ejecta that clearly set them apart from ordinary SNe Ic (non-relativistic). 
The thermal properties of relativistic SNe are also analogous to the very energetic,
fast expanding SNe associated with sub-E GRBs and ordinary GRBs 
(\citealt{Soderberg10}, M14). 
\emph{However, a distinctive property of relativistic SNe is their
significantly fainter X-ray emission (Fig. \ref{Fig:Xray}) that implies the lack of a luminous X-ray 
component arising from the central engine activity at late times ($\delta t\sim 20\,\rm{d}$,
Fig. \ref{Fig:Radio2Xray}).} 
With $E_{\rm{\gamma,iso}}<10^{47}\,\rm{erg}$, the prompt 
$\gamma$-ray energy released by the relativistic SN\,2012ap is also considerably below 
the level of sub-E GRBs (Fig. \ref{Fig:LxEiso}).\footnote{For the other relativistic SN\,2009bb the
observational limit on the promptly released  $E_{\rm{\gamma,iso}}$ is unfortunately not as 
constraining (Fig. \ref{Fig:LxEiso}).} The higher metallicity of the environment and the conspicuous
presence of helium in the ejecta of the two known relativistic SNe also sets them apart from 
sub-E GRBs and GRBs.

These findings call for some crucial diversity in the properties of the engines 
and/or of the progenitors of relativistic SNe and sub-E GRBs (and ordinary GRBs as well). 
We showed that the observations of relativistic SNe are consistent with the picture of 
 jet-driven explosions where the jet just barely fails to breakout from the progenitor
star. This scenario naturally explains (i) the lack of evidence for central engine activity at late times and
(ii) the deep limit to the promptly released $\gamma$-ray energy.

The failed jet breakout might be due to an intrinsically short-lived engine (but same progenitor properties)
or to a different progenitor structure between relativistic SNe and sub-E GRBs. At the time of
writing, with only two relativistic SNe discovered so far (through targeted optical surveys), 
observations do not allow us to distinguish between these two scenarios. 
A significantly larger sample of relativistic SNe found through \emph{untargeted} SN searches in the
optical and radio band is clearly needed to 
deeply understand their connection to GRBs, 
build a complete picture of E-SNe and constrain which unique property
differentiates failed breakouts from successful, fully relativistic jets.
The Large Synoptic Survey Telescope (LSST, \citealt{Ivezic08})
 is expected to discover hundreds of SNe Ic every year, thus in principle providing the significantly
larger sample of E-SNe that is needed to deeply understand the connection between relativistic SNe
and GRBs.  However,  as we demonstrate here,  coordinated radio \emph{and X-ray} follow up 
is essential to identify E-SNe from their ordinary
counterparts and determine the properties of the engines that power their explosion.

\acknowledgments 
We thank the referee for useful comments and suggestions that improved the quality of
our  paper. R.M. is grateful to the Aspen Center for Physics and the NSF Grant \#1066293 for hospitality 
during the completion of this work. Support for this work was provided by the David and Lucile Packard Foundation 
Fellowship for Science and Engineering awarded to A.~M.~S. 
\bibliographystyle{apj}

\end{document}